# Observation of Acoustically Induced Dressed States of Rare-Earth Ions

Ryuichi Ohta[1], Grégoire Lelu[1], Xuejun Xu[1], Tomohiro Inaba[1], Kenichi Hitachi[1], Yoshitaka Taniyasu[1], Haruki Sanada[1], Atsushi Ishizawa[2], Takehiko Tawara[3], Katsuya Oguri[1], Hiroshi Yamaguchi[1], and Hajime Okamoto[1]

[1]NTT Basic Research Laboratories, NTT Corporation, 3-1 Morinosato Wakamiya, Atsugi, Kanagawa 243-0198, Japan
[2]College of Industrial Technologies, Nihon University, 1-2-1 Izumi, Narashino, Chiba 275-8575, Japan
[3]College of Engineering, Nihon University, 1 Tokusada Nakagawara, Tamura, Kouriyama, Fukushima, 963-8642, Japan

Acoustically induced dressed states of long-lived erbium ions in a crystal are demonstrated. These states are formed by rapid modulation of two-level systems via strain induced by surface acoustic waves whose frequencies exceed the optical linewidth of the ion ensemble. Multiple sidebands and the reduction of their intensities appearing near the surface are evidence of a strong interaction between the acoustic waves and the ions. This development allows for on-chip control of long-lived ions and paves the way to highly coherent hybrid quantum systems with telecom photons, acoustic phonons, and electrons.

Interactions between two-level systems (TLSs) and external fields are of fundamental interest in various fields of physics and quantum technologies. Such interactions allow the TLSs to be controlled through external fields, enabling optical and electrical excitations and readout of the TLS states. Under a strong field, the TLSs are renormalized by the field, leading to hybrid states called "dressed states". Such dressed states are a universal concept across many physical platforms [1, 2] and can be used for controlling and engineering TLSs, such as through coherent state manipulation [3, 4], population transfer between different spin states [5, 6], and extension of the coherence time of the systems [7].

Dressed states generated by rapid modulation of TLSs through an acoustic field are particularly interesting as they offer the possibility of local manipulation of quantum devices. Because the acoustic fields can be spatially confined within the order of their wavelengths, typically several micrometers in the vertical or horizontal direction, this scheme is advantageous for enhancing the interaction between the field and the TLSs and also for device integration. So far, acoustically induced dressed states have been used to couple excited electrons with different spin configurations. This has a potential application to continuous dynamic decoupling by acoustic phonons and may lead to robust two-level systems in solid platforms [6]. Moreover, these states would be useful in the fields of quantum acoustics [8] and optomechanics [9], as a way to build hybrid quantum systems with photons, electrons, and phonons. However, the ability to generate optically accessible dressed states, which are evidenced by the sideband peaks in the optical spectra, has so far been limited to TLSs with short decay times (< 100 ns), such as in quantum dots [10-13] and diamond color centers [6, 14, 15]. These short decay times hinder the usefulness of the dressed states especially in quantum memories and repeaters which require long coherence times (> 1 ms). If optically accessible acoustically induced dressed states could be realized in long-lived TLSs, it would enable great control over quantum states and be a breakthrough for highly coherent hybrid quantum systems.

Rare-earth ions in crystals are attractive TLSs whose resonances cover both optical and microwave frequencies with extremely long coherence times. They have been incorporated in various solid-state quantum devices, such as quantum memories and repeaters [16-18], spin qubits [19, 20], and transducers operating between optical and microwaves photons [21–23], in which the optical properties are externally manipulated with magnetic [24, 25] and electric fields [26, 27]. Acoustic modulation of rare-earth ions has also been demonstrated in an erbium-ion-embedded micromechanical resonator by using mechanical resonance at 1 MHz [28]. However, acoustically induced dressed states have so far remained unrealized because the mechanical frequencies have been too low to modulate the TLSs in the resolved sideband regime.

Here, we demonstrate optically accessible dressed states that are acoustically induced by the interaction of erbium (Er) ions and a standing surface acoustic wave (SAW). In this system, the acoustic resonance frequency (2.13 GHz) exceeds the optical linewidth of Er ions (500 MHz); thus, this system meets the resolved sideband condition. The rapid modulation of the optical resonance in this regime leads to the appearance of multiple sidebands in the photoluminescence excitation (PLE) spectra, which correspond to the acoustically induced dressed states of the Er ions (Fig. 1(a)). This feature can be reproduced with a model based on frequency modulation of an ensemble of two-level systems. We found that the maximum modulation index, which is proportional to the coupling strength between electrons and phonons, reaches 5.0 near the crystal surface. This large modulation index allows the destructive interference and reduction of the PL intensities induced by the SAW. It indicates the suppression of the direct electron-photon coupling and thus enables one to control the excitation of Er ions via acoustic waves. These acoustically induced dressed states would be useful in the fields of quantum acoustics and optomechanics and may pave the way to on-chip hybridization of the telecom photons, long-lived electrons, and acoustic phonons.

Figures 1(b) and 1(c) show schematic and microscope images of the SAW device fabricated on a $^{170}$Er: yttrium orthosilicate (YSO) crystal with an Er density of 50 ppm. We fabricated 800-nm-pitch interdigital transducers (IDTs) and Bragg reflectors (BRs) on YSO crystal with a 200-nm-thick aluminum nitride (AlN) piezoelectric layer [29]. Figures 1(d) shows the numerically calculated shear strain ($\varepsilon_{xz}$) induced by a SAW with the finite element method. It shows that the strain field of the SAW penetrates the substrate with a decay constant ($d_m$) of 3.5 μm from the boundary of the AlN and YSO (see also Fig. 4(b)), while the strain field is widely distributed in the x-y plane between the two BRs. The wavelength of the SAW was designed to be 1.6 μm, which corresponds to a mechanical resonance frequency of 2.1 GHz [29].

Figure 2(a) shows the experimental setup for the optical spectroscopy (red and yellow arrows) and frequency response measurements (green arrows). For the optical spectroscopy, the pump laser frequency was stabilized with an optical frequency comb laser near the resonance frequency of the $Y_1$-$Z_1$ transition of Er ions (1536.4 nm) and finely tuned with an electro-optic modulator [30]. To reduce the peak broadening induced by the heating effect due to the carrier and SAW excitation, the pulsed optical pump and acoustic drive were injected with duty ratios of 0.8% and 1.6%, respectively. The pump laser was focused on the surface (3 μm waist) between the two IDTs. The PL intensities of the $Y_1$-$Z_2$ transition (1546.4 nm) were measured with the detectors. For the frequency response measurements, the SAW was continuously excited with a drive voltage of 12 $V_{rms}$, while the frequency was swept near 2.1 GHz. The frequency response was measured with a Sagnac optical interferometer and a vector network analyzer. The sample was measured at 10 K in a low-pressure chamber (< 10$^{-4}$ Pa).

The measured mechanical response of the SAW device is shown in Fig. 2(b). The $Q$ factor of the mechanical resonance at 2.13 GHz (= $\omega_m/2\pi$), as estimated from a Lorentzian fitting, was 400. The frequency response of the electrical transmission between the two IDTs was also measured. It showed acoustic resonance at 2.13 GHz [29]. Figure 2(c) shows the PLE spectrum of Er ions without a SAW excitation. The full width at half maximum ($\Gamma/2\pi$) of the optical linewidth was 500 MHz, which is lower than the acoustic resonance frequency, thereby allowing for an acoustic modulation of the ion ensemble in the resolved-sideband regime ($\omega_m > \Gamma$). Figure 2(d) shows the time-resolved PLE intensities without SAW excitation. It confirms that the relaxation lifetime is extremely long ($\tau = 10.1$ ms), which is a remarkable aspect of the Er ions.

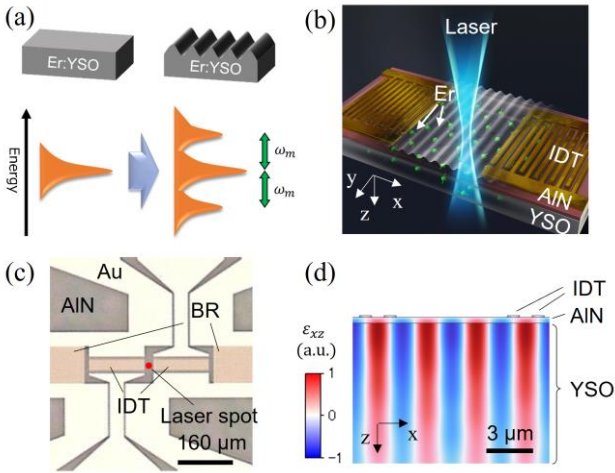

Figure 1: (a) Schematic image of the generation of acoustically induced dressed states of an Er ion ensemble. $\omega_m$ is the angular frequency of the SAW. (b) Schematic illustration of the SAW device fabricated on an $^{170}$Er:YSO crystal. Laser is focused at the center of two IDT electrodes. (c) Microscope image of the fabricated device. Red circle indicates the point of the laser irradiation. (d) Numerically calculated shear strain induced by SAW in the x-z cross-section.

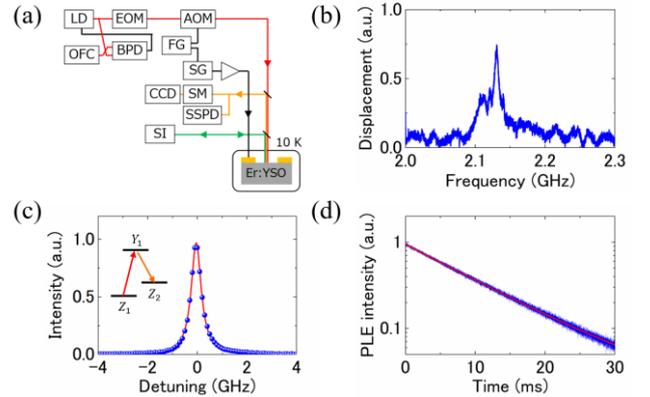

Figure 2: (a) Experimental setup. LD: laser diode, EOM: electro-optic modulator, AOM: acousto-optic modulator, OFC: optical frequency comb, BPD: balanced photo detector, FG: function generator, SG: signal generator, CCD: charge coupled device, SM: spectrometer, SSPD: superconducting single photon detector, and SI: Sagnac optical interferometer. (b) Frequency response of the displacement. (c) PLE spectrum without acoustic drive. The inset shows the energy levels used for the optical pump and luminescence measurements. (d) Optical decay of the excited state of the Er ions measured through the PLE intensity.

To observe the acoustically induced dressed states, we combined the PLE measurement with the excitation of a SAW. Figure 3(a) shows the PLE spectra taken under acoustic driving ($V_d$) with 19.2 $V_{rms}$ and a driving frequency ($f_d$) of 2.13 GHz (blue plots). For comparison, the PLE spectra without acoustic driving is plotted in gray. In addition to the center peak, which corresponds to the direct optical transition, sideband peaks up to $n = 3$ appear at frequencies of $n \times f_d$, where $n$ is an integer. These sideband peaks correspond to the acoustically induced dressed states of the Er ions, which are the renormalized states of the ions under the strong acoustic field.

To quantitatively analyze the dressed states, we calculated the spectra with a model including time-varying strain that modulates the transition frequency of the Er ion ensemble. In the case of a single ion, the PLE spectrum of the dressed state ($I_{sing}$) is described by multiple peaks weighted with an $n$-th Bessel function of the first kind ($J_n[\chi]$; Eq. (1a)) in the same manner as a quantum dot [10] or nitrogen vacancy center [6].

$$I_{sing}(\omega) \propto \sum_n \frac{J_n^2[\chi] \times \Gamma}{(\omega - \omega_{Er} - n\omega_m)^2 + \Gamma^2} \quad (1a)$$

$$\chi = G_V \times \frac{V_d}{\omega_m} \quad (1b)$$

In these systems, the modulation index, called the Bessel parameter, ($\chi$) is defined by the product of the modulation coefficient of the resonance frequency of the ions ($G_V = d\omega/dV$) and the input voltage normalized by the acoustic resonance frequency $V_d/\omega_m$, where $G_V$ corresponds to the overall efficiency of generating the acoustically induced dressed states in this device (Eq. (1b)). In the case of the ion ensemble, where the ions are uniformly distributed in the crystal, these modulation effects should accumulate through all ions within the area of the laser spot. As shown in Fig. 1(d) and Fig. 4(b), the strain field of the SAW, which is proportional to $\chi$, exponentially decreases along the z-axis with $d_m$ of 3.5 μm. On the other hand, the optical field ($E_{opt}$) is distributed according to a Gaussian beam profile within a focal depth ($d_{opt}$) of 4.8 μm (see Fig. 4(b)), which was estimated by fitting the PLE spectra [29].

The observed PLE spectra include responses from all of the optically excited ions distributed in the z direction ($I_{ensem}(\omega)$) and are well fitted with the following equations:

$$I_{ensem}(\omega) \propto \int E_{opt}(z) \times \sum_n \frac{J_n^2[\chi(z)] \times \Gamma(z)}{(\omega - \omega_{Er}(z) - n\omega_m)^2 + \Gamma(z)^2} dz \quad (2a)$$

$$E_{opt}(z) = E_0 d_{opt}/\sqrt{z^2 + d_{opt}^2} \quad (2b)$$

$$\chi(z) = \chi_0 \exp(-z/d_m) \quad (2c)$$

$$\omega_{Er}(z) = \omega_0 + \Delta\omega + \delta\omega \exp(-z/d_m) \quad (2d)$$

$$\Gamma(z) = \Gamma_0 + \Delta\Gamma + \delta\Gamma \exp(-z/d_m) \quad (2e)$$

where $E_0$ is the normalized optical intensity, $\chi_0$ is the maximum Bessel parameter obtained at the boundary of the AlN and YSO, $\omega_0$ ($\Gamma_0$) is the intrinsic resonance frequency (linewidth) measured without SAW driving. $\Delta\omega$ ($\Delta\Gamma$) is the z-independent frequency shift (linewidth broadening), and $\delta\omega$ ($\delta\Gamma$) is the coefficient of the frequency shift (linewidth broadening) that exponentially depends on the z position. $\Delta\omega$ ($\Delta\Gamma$) shifts (broadens) all peaks in equal, whose origin we consider is the heating effect coming from the dissipation of SAW energy [6, 14, 15]. On the other hand, the last term of Eq. 2d (2e) dominantly shifts (broadens) the higher order sidebands, which originate from the luminescence from the strained ions located near the surface (i.e., at small z) [31, 32]. The last term of Eq. 2d leads to a difference in the frequency shift between the center peak and the sidebands. This causes the asymmetry in amplitude between the blue and red sidebands shown in Fig. 3(b) and 3(c) [29]. We used Eq. (2a-2e) to fit the PLE spectra for various drive voltages, where the spectra for $V_d$ = 19.2, 7.6, and 3.0 $V_{rms}$ are shown in Fig. 3(b)-3(d). Good agreement between the experimentally observed spectra and the fitted curves confirm that our model describes the overall behavior of the acoustically induced dressed states.

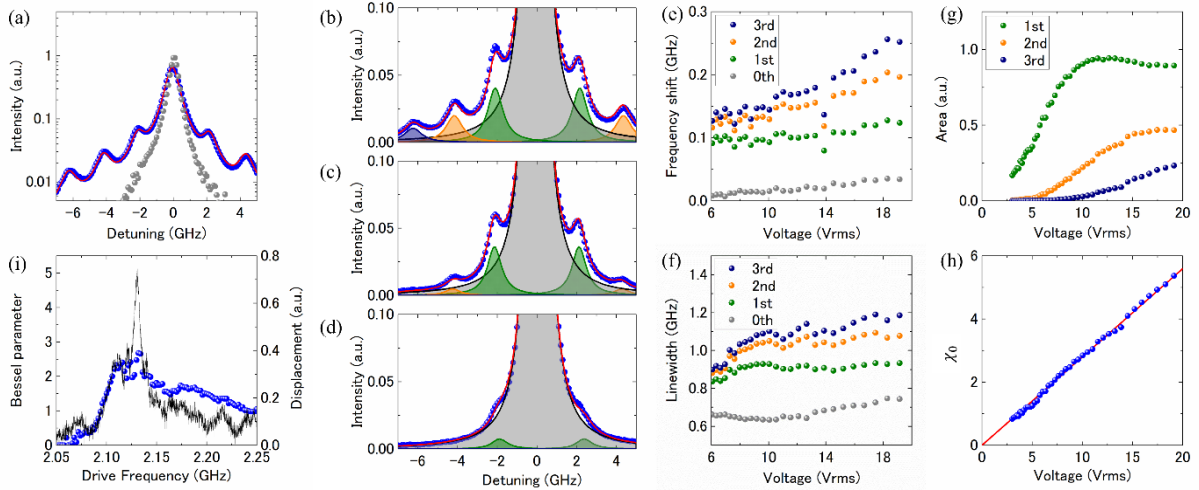

Figure. 3: (a) PLE spectra with (blue circles) and without (gray circles) acoustic driving of 19.2 $V_{rms}$. Red lines are fittings given by Eq. 2. (b-d) PLE spectra for drive voltages of 19.2, 7.6, and 3.0 $V_{rms}$. We decomposed the spectra into the 0th (gray), 1st (green), 2nd (orange), and 3rd (purple) sidebands by using Eq. 2a-e. (e,f) Drive voltage dependence of the frequency shift (e) and linewidth (f) of the center peak (gray circles), 1st (green circles), 2nd (orange circles), and 3rd (purple circles) sidebands. (g) Drive voltage dependence of the peak area of the 1st (green circles), 2nd (orange circles), and 3rd (purple circles) sidebands. (h) Drive voltage dependence of $\chi_0$. The red line is the linear fitting curve. (i) Drive frequency dependence of the Bessel parameters determined from the sideband peaks (circles) at a drive voltage of 12 $V_{rms}$. The solid line shows the frequency response of the mechanical displacement measured with a Sagnac interferometer.

Figures 3(e) and 3(f) show the frequency and linewidth of the center peak and each sideband with respect to the driving voltage, which were estimated by fitting the experimental data with Eq. 2. They show that the frequency shift and linewidth broadening of the higher order sidebands are larger than that in the lower order sidebands or the center peak. This suggests that SAW strain, which is larger in the regime near the surface, is the main cause of these changes, because the higher order sidebands are dominated by excited ions located near the surface (at small z). Figures 3(g) and 3(h) show the drive-voltage dependence of the peak area in each sideband and $\chi_0$. The linear dependence of $\chi_0$ on the drive voltage leads to $G_V$ = 3.34 $GHz/V_{rms}$, while $\chi_0$ exceeds 5.0 with high drive voltages. We also found that the generation of the sidebands strongly depends on the SAW drive frequency (Fig. 3(i)). The matching of the overall tendency of $\chi_0$ with the mechanical displacement indicates that the dressed states were generated by the strain field induced by the SAW [29].

By taking into account the strain distribution along the z-axis (Fig. 1(d) and Fig. 4(b)), we can estimate the z-position dependence of the acoustically induced dressed states and the sideband amplitude [29]. Here, we should note that the optical field depth is deeper than the strain depth, i.e., $d_{opt} > d_m$ (Fig. 4(b)). Therefore, the experimentally obtained PLE spectra include the signal from the ions with a different strain-field strength. We decomposed the spectrum (obtained with the drive voltage of 19.2 $V_{rms}$) with respect to the z-position based on Eq. 2, as shown in Fig. 4(a). The decompositions indicate that each sideband originates from ions located at different depths and the heights of the sidebands become higher than the center peak within 8 μm from the top surface. Owing to the large $\chi_0$ near the surface, maximum and minimum points both appear in the intensities of the 0th, 1st, and 2nd sidebands. The minimum points, where $J_n[\chi(z)] = 0$ (points A, B, and B' in Fig. 4(a)), are particularly interesting. They are caused by destructive interference from the cascaded sidebands [2]. For example, the first sidebands interfere with the center peak via their cascaded sidebands destructively, leading to a minimization of the optical intensity at point A in Fig. 4(a). The appearance of the interference reveals that the electron states are coherently modulated by the acoustic strain, so that the energy transduction is a coherent process. The annihilation of the center peak at point A indicates that we can externally manipulate the effective dipole moment of the electrons and set the excited ions to be optically dark. Such dark states could be used to extend the relaxation rate of the excited ions and thus would be useful for memory applications.

In the current device, a signature of the destructive interference is the saturation and the small reduction in the peak area of the first sidebands shown in Fig. 3(g). On the other hand, a reduction in amplitude to zero cannot be confirmed, because the observed intensity is the sum of the intensities along the z direction, which include the luminescence from weakly strained or unstrained ions. In order to utilize the destructive interference, one should optically excite only the ions around the top surface, in which the acoustic strain is large. In this respect, it would be effective to incorporate nanophotonic waveguides and cavities into Er: YSO crystals [33], which would give an evanescent optical field within 1 μm from the surface. The use of epitaxially grown thin films containing Er ions would be another way to localize ions within 100 nm of the surface [34]. The combination of nanophotonic and acoustic devices on epitaxially grown Er-doped substrates would substantially improve the controllability of the ions with this acoustically induced dressed architecture.

In summary, this letter examined the acoustically induced dressed states of Er ions in a YSO crystal. The ability to generate an acoustically induced dressed state is not only important for engineering long-lived optical states; it is also important in terms of quantum acoustics and optomechanics, once the acoustic field is quantized. By using phonon sidebands, coherent energy transduction between photons, electrons, and phonons becomes possible with appropriate frequency matching of the laser, ions, and acoustic waves. For instance, Rabi oscillations (entanglement) between excited electrons and acoustic phonons are driven by absorption of photons whose frequencies are red- (blue-) detuned from the resonance of the ions. Owing to the long-lived nature of the rare-earth ions, the decay times of the excited electrons, which directly interact with the photons, are much longer than those of the phonons. Thus, we can construct an optomechanical system in the so-called reversed dissipation regime [28, 35] with optical photons. In this regime, multiple acoustic pulses can be applied to the ions within their decay times. This would allow us to coherently control the excited electrons by using acoustic phonons and devise acoustic pump-probe spectroscopy and phonon-driven echo measurements in an analogous way to optical pulses. Moreover, the dressed states reflect the physical characteristics of Er ions, such as spin-selection rule in the transitions, magnetic field dependence on energy levels with various g-factors, and nonlinear responses due to the saturations of excitations. The acoustic modulation of these properties develops the fundamental study of solid-state physics in rare-earth ions. The resonance frequencies of the Er ions are at telecom wavelengths, so such devices can be easily interconnected with optical fibers. The presented results will enable on-chip control of highly coherent light-matter interfaces and will advance the field of quantum acoustics and optomechanics through the development of integrated hybrid quantum architectures for large-scale quantum networks.

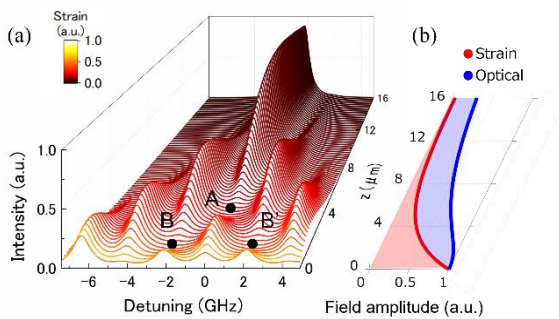

Figure. 4: (a) Decomposition of the PLE spectrum at $V_d$ of 19.2 $V_{rms}$ with respect to the depth of the YSO crystal. Color corresponds to the normalized strain. A, B, and B' indicate the points where the intensity becomes zero due to the destructive interference of the excitation. (b) Strain and optical field distribution along the z-axis.

We thank S. Sasaki, D. Hatanaka, M. Kurosu, M. Hiraishi, and S. Yasui for technical support and discussions. This work was supported by JSPS KAKENHI, Grant Number JP23H01112.

# Supplemental Materials for
# "Observation of Acoustically Induced Dressed States of Rare-Earth Ions"

## DESIGN AND FABRICATION OF THE SAMPLE

In this section, we describe the procedure of fabricating the SAW device. First, we sputtered 200-nm-thick AlN thin films at 800 °C on $^{170}$Er:YSO crystals with an ion concentration of 50 ppm (*Scientific Materials*). The C-axes of AlN were oriented along the b-axis of the crystal (z-axis of the device). The X-ray rocking-curve (XRC) linewidth of the sputtered AlN was 4.0 degrees. Next, a 20-nm-thick gold layer was deposited on the AlN to form interdigital transducers (IDTs) with an 800-nm-pitch and 100 periods. 200-period-Bragg reflectors (BRs) with the same pitch as the IDT were set outside the IDTs to form a Fabry-Pérot phonon cavity to enhance the strain field with acoustic resonances. The propagation direction of the SAW was along the D2 direction of the crystal (x-axis of the device). The length of the IDTs and the distance between them were 30 and 25 µm, respectively.

We designed the pitch of the IDTs to stand the SAW around 2 GHz. A lower frequency would improve the spatial overlap between the optical and strain fields because of the larger penetration depth of the strain ($d_m$). However, to clearly observe the sidebands as separable peaks, we chose a SAW frequency around 2 GHz.

Figure S1 shows the strain distribution of $\varepsilon_{xx}$, $\varepsilon_{zz}$, and $\varepsilon_{xz}$ calculated with a finite element method (FEM). In this two-layer structure of AlN and YSO, $\varepsilon_{xx}$ and $\varepsilon_{zz}$ do not penetrate in the YSO, while $\varepsilon_{xz}$ penetrates the YSO with a decay constant of 3.5 µm. Thus, $\varepsilon_{xz}$ is the major component with which to make the dressed states and the other components are negligible.

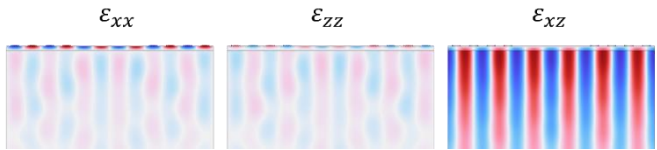

Fig. S1: Distribution of each strain component simulated by FEM calculations.

## ELECTRICAL TRANSMISSION SPECTRUM OF THE SAW DEVICE

We measured the electrical transmission spectrum ($S_{21}$) between the two IDT electrodes at 10 K. Figure S2(a) shows the $S_{21}$ spectrum, whose y-axis is a linear scale of the transmitted power. Figure S2(b) shows the square of the mechanical amplitude (spectral power density) measured by the Sagnac interferometer. In both cases, a mechanical resonance at 2.13 GHz appears. The Q factor derived from $S_{21}$ is about 500, which is similar to that given by the Sagnac interferometer. A resonance at 2.12 GHz also appears in the S21 spectrum (Fig. S2(a)), while one is not clearly observable in the optically measured spectrum (Fig. S2(b)). The resonance only found in the $S_{21}$ spectrum at 2.12 GHz might be caused by the reflection between the IDT and the Bragg reflector, whose frequency is slightly different from the resonance of the cavity. We used the acoustic resonance at 2.13 GHz for the experiments in the main text.

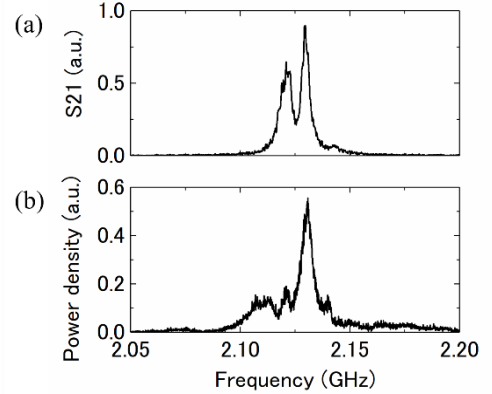

Fig. S2: (a) Electrically measured transmission spectrum ($S_{21}$) between two IDT electrodes. (b) Spectral power density of SAW measured by optical interferometry.

## DRIVE FREQUENCY DEPENDENCE AT LOWER VOLTAGE AND THE ORIGIN OF THE SIDEBANDS

The general tendencies of the Bessel parameter estimated from the PLE spectra and the optically measured mechanical displacement for the driving frequency are similar (Fig. 3(i)). However, they do not perfectly match at the resonance of 2.13 GHz and frequencies above it. We cannot precisely explain this discrepancy, but one reason may be the different driving schemes for the Sagnac interferometry (continuous driving) and the sideband observation (pulse driving), as the agreement is better for the smaller drive voltage of 4.2 $V_{rms}$ in Fig. S3(a).

Although a small discrepancy remains, the overall tendencies of the two measurements are similar. For instance, driving at a lower frequency (<2.07 GHz) does not create dressed states, which excludes microwaves coming from the IDTs as a potential origin. The PLE spectra with the resonant drive (2.13 GHz) was distinctly different from that with the slightly detuned drive (for instance, 2.15 GHz), as shown in Fig. S3(b). The difference between the Bessel parameters estimated for these two PLE spectra is not as large as that in the mechanical displacement measured by Sagnac interferometry, but the results suggest that the acoustic resonance on the dressed states makes a significant contribution. Thus, we consider that the dressed states were caused by SAW strain.

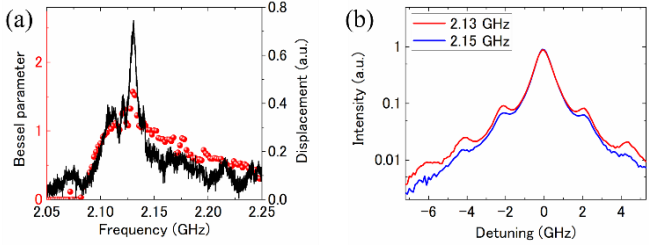

Fig. S3: (a) Drive frequency dependence of the Bessel parameter for pulse driving with $V_d = 4.2$ $V_{rms}$ (red plots). The solid line shows the frequency response of the mechanical displacement measured by continuous driving with $V_d = 12$ $V_{rms}$ with a Sagnac interferometer. (b) PLE spectra taken with drive frequencies of 2.13 and 2.15 GHz and $V_d$ of 12 $V_{rms}$.

## ESTIMATION OF THE OPTICAL DECAY LENGTH

To fit the PLE spectra shown in Fig. 3(a)-3(d), we set $d_m$ and $d_{opt}$ to 3.5 and 4.8 μm, respectively, and took $E_0$, $\Delta\omega$, $\delta\omega$, $\Delta\Gamma$, and $\delta\Gamma$ to be fitting parameters. Here, $d_m$ was derived to be 3.5 μm from the FEM simulation. $d_{opt}$ was 4.8 μm; this value gives the best fitting for the largest drive voltage of 19.2 $V_{rms}$, where a larger drive voltage should lead to a more reliable $d_{opt}$ value because of the appearance of a large number of sidebands originating from ions located at different depths.

## DECOMPOSED SPECTRA AND DESTRUCTIVE INTERFERENCE

The decomposed spectrum at each depth (Fig. 4(a)) was derived by fitting the experimental data (shown in Fig. 3(a)) to the form of Eqs. 2a-e in a minute integration range of z as follows,

$$I(z) = \int_z^{z+\Delta z} E_{opt}(z) \times \sum_n \frac{J_n^2[\chi(z)] \times \Gamma(z)}{(\omega - \omega_{Er}(z) - n\omega_m)^2 + \Gamma(z)^2} dz \quad (S1)$$

where $\Delta z$ = 200 nm.

In order to reduce the excitation probabilities to zero, the maximum modulation index should be higher than 2.5. This can be understood by examining the graph that shows the relation between $J_n^2[\chi]$ and the Bessel parameter (Fig. S4). When the modulation index is 2.5, the amplitude of the 0th-order Bessel function is zero (see the red plots in Fig. S4). This corresponds to point A in Fig. 4(a), where the direct transition probability of the ions becomes zero. The second points appear when the modulation index is 3.8 (see the green plots in Fig. S4). This corresponds to point B and B' in Fig. 4(a), where the amplitude of the 1st–order sidebands become zero.

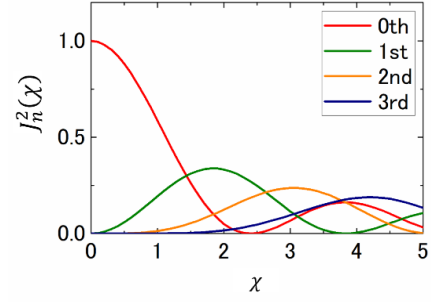

Fig. S4: Relation between the Bessel parameter, $\chi$, and the square of the amplitude of the $n$-th order Bessel function.

## FREQUENCY SHIFT AND LINEWIDTH BROADENING CAUSED BY THE HEAT OF THE SAW

The theoretical form, Eq. 2, fits the experimental data shown in Fig. 3(b-d). Here, the z-independent term including $\Delta\omega$ ($\Delta\Gamma$) in Eq. 2d (2e) shifts (broadens) all peaks equally, whose origin we consider is not strain but rather the heating effect coming from the dissipation of SAW energy. The heating effect can be evaluated by comparing the experimentally measured temperature dependence of the frequency shift and linewidth broadening with $\Delta\omega$ and $\Delta\Gamma$ estimated by the fitting. Figures S5(a) and S5(b) show the temperature dependence of the frequency shift and linewidth broadening of the Er resonance measured at 10-20 K in a similar sample of Er-doped YSO crystal, where a heater in the cryostat was used to increase the sample temperature. Figure S5(c) and S5(d) show $\Delta\omega/2\pi$ and $\Delta\Gamma/2\pi$ estimated by making a fitting for different drive voltages in the present SAW device. They give $\Delta\omega/2\pi$ < 30 MHz and $\Delta\Gamma/2\pi$ < 200 MHz for $V_d \leq 20$ $V_{rms}$. A comparison of these values (Figs. S5(c) and S5(d)) with the measured temperature dependence of the frequency shift and linewidth broadening (Figs. S5(a) and S5(b)) suggests that the temperature increase for $V_d \leq 20$ $V_{rms}$ is within 4 K. We consider that this heating effect can be further reduced by optimizing the device design, for example, by reducing the IDT width.

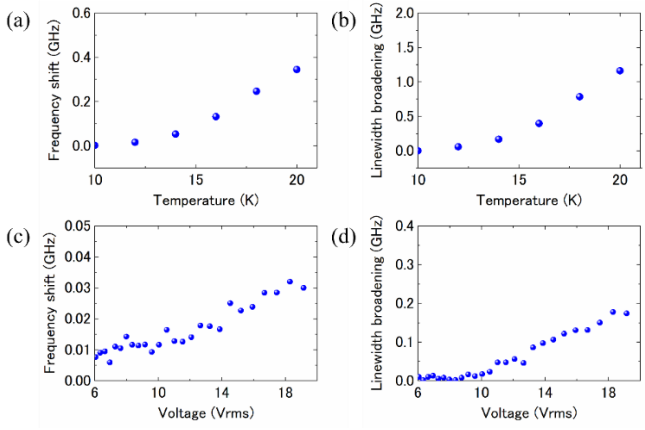

Fig. S5: Temperature dependences of the frequency shift (a) and linewidth broadening (b) of the optical resonance measured in an Er-doped YSO crystal without drive voltages. (c-d) Those caused by SAWs, which correspond to $\Delta\omega/2\pi$ and $\Delta\Gamma/2\pi$ in Eqs. 2d and 2e.